\documentclass{appolb}
\usepackage{epsfig}

\preprint{}

\begin{document}

\title{Explanation of the RHIC $p_\perp$-Spectra \\
in a Single-Freeze-Out Model
\thanks{Supported by the Polish State Committee for
Scientific Research, grant 2 P03B 09419.}}
\author{Wojciech Florkowski and Wojciech Broniowski 
\address{The H. Niewodnicza\'nski Institute of Nuclear Physics\\
PL-31342 Krak\'ow, Poland}
}
\maketitle

\begin{abstract}
The $p_\perp$-spectra of hadrons measured at RHIC are very well 
described in a model which assumes that the chemical and thermal
freeze-outs occur simultaneously. The model calculation includes
all hadronic resonances and uses a simple parametrization of the
freeze-out hypersurface. 
\end{abstract}

\bigskip

In this talk we present a simple model describing the $p_\perp$-spectra 
of various hadrons measured recently at RHIC. Our approach
combines the thermal model, used frequently in the studies of the
relative hadron yields \hbox{[1-10]},
with a  model of the hydrodynamic expansion of 
matter at freeze-out. The main assumptions of the model are as 
follows \hbox{[11-13]}:
First of all, we assume that the chemical freeze-out and the thermal
freeze-out occur at the same time. This means that we neglect elastic
rescattering after the chemical freeze-out.  Secondly, we include all
hadronic resonances in both the calculation of the hadron
multiplicities and the spectra. In particular, all cascade decays are
taken into account exactly in a semi-analytic fashion. Finally, we
assume a simple form of the freeze-out hypersurface, which is a
generalization of the Bjorken model \cite{bjorken} (see also
\hbox{[15-20]})
%
\begin{equation}
\tau = \sqrt{t^2-x^2-y^2-z^2} = \hbox{const}.
\label{tau}
\end{equation}
The hydrodynamic flow on the freeze-out hypersurface (\ref{tau})
is taken in the form resembling the Hubble law
\begin{equation}
u^\mu = {x^\mu \over \tau} = {t \over \tau} \left(1, {x \over t},
{y \over t},{z \over t}\right).
\label{umu}
\end{equation}

\begin{figure}[t]
\epsfysize=10.75cm
\mbox{\hspace{-1.5cm} \epsfbox{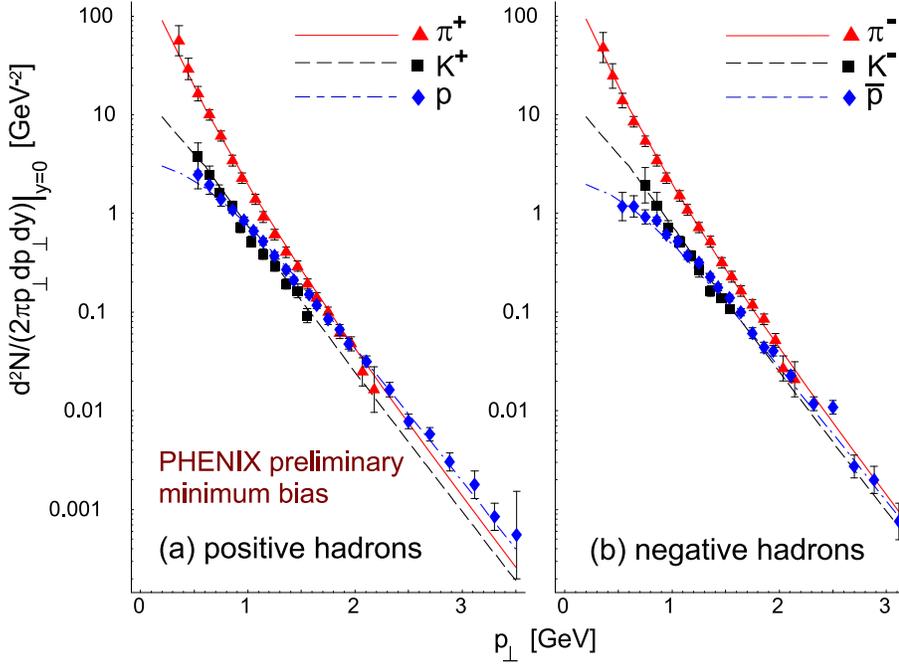}}
\label{minb}
\caption{The $p_\perp$-spectra at midrapidity of pions (solid line),
kaons (dashed line) and protons or antiprotons (dashed-dotted line).
The model calculation is compared to the PHENIX preliminary minimum-bias
data \cite{phenix}, Au + Au collisions at $\sqrt{s}$ = 130 GeV A. }
\label{fig1}
\end{figure}

Recently, several arguments have been accumulated in favor of our
first assumption. The measurements of the transverse HBT radii at RHIC
show that the hadronic system does not evolve for a long time.  The
ratio $R_{out}/R_{side}$ is close to unity in the whole range of the
measured transverse momentum (for a compilation of the recent HBT
measurements see, e.g., Ref. \cite{phenixhbt}), indicating that the
pion emission time is short in comparison with a typical transverse
size of the system.  The assumption about the single freeze-out solves
also the antibaryon puzzle \cite{rapp}. Since the annihilation cross
section for $p {\bar p}$ pairs is much larger than the elastic cross
section, most of the protons would annihilate with antiprotons during
the long way from the chemical to the thermal freeze-out. Such effect
is not seen. In addition, let us mention that a rapid transverse
expansion of the firecylinder, as found at RHIC also in our model,
suppresses the collision rate and makes the potential gap between the
two freeze-outs smaller.

Our model has two thermodynamic and two geometric (expansion)
parameters.  The two thermodynamic parameters are obtained from the
analysis of the ratios of the hadron multiplicities measured at RHIC.
The fit to 9 independent ratios yields the values of the temperature
and the baryon chemical potential: $T$ = 165 MeV, $\mu_B$ = 41 MeV
\cite{fbm}. Since the particle ratios depend weakly on the centrality
of the collision \cite{qm2001}, the thermodynamic parameters may be
regarded as the universal parameters (independent of centrality).  The
two geometric parameters are $\tau$ and $\rho_{\rm max}$.  The
parameter $\rho_{\rm max}$ determines the transverse size of the
firecylinder at the freeze-out,
\begin{equation}
\rho = \sqrt{x^2 + y^2} \le \rho_{\rm max}.
\label{rhomax}
\end{equation}
Clearly, the values of $\tau$ and $\rho_{\rm max}$ depend on the
considered centrality class of events. For the minimum-bias data,
which average over centralities, we find: $\tau = 5.55 \hbox{ fm}$ and
$\rho_{\max} = 4.50 \hbox{ fm}$, whereas for the most central
collisions we find: $\tau = 7.66 \hbox{ fm}$ and $\rho_{\max} = 6.69
\hbox{ fm}$ \cite{wbwf}. The calculation of the spectra (and
determination of the geometric parameters) is based on the standard
Cooper-Frye formalism. The details of our method, especially of the
technical problems concerning the treatment of the resonances, are
given in the Appendix of Ref. \cite{str}.

In Fig.  1 we compare the model predictions for the $p_\perp$-spectra
of pions, kaons and (anti)protons with the PHENIX minimum bias
preliminary data \cite{phenix} (the official data released recently
\cite{phenixoff} agree with those shown on the plot).  The quality of
the fit is very good. The model curves cross practically all the data
points within error bars. Even the high-$p_\perp$ data are
reproduced. Also, the convex shape of the pion spectra is provided by
the model. We have checked that this is a consequence of the radial
flow \cite{wbwf}. The main effect coming from the resonance decays is
an effective cooling of the spectrum by 30-40 MeV -- the inverse slope
of the spectrum becomes smaller as discussed in more detail in
Ref. \cite{fbm}.  The effect of the cooling of the spectrum by the
decays of the resonances explains a difference between the high
temperature of the chemical freeze-out (in different approaches it
turns out to be very similar and close to 170 MeV) and a smaller
``apparent'' temperature (inverse slope) inferred from the analysis of
the spectra, which is typically not more than 130 MeV.

In Fig. 2 we show our results for the most central collisions.  In the
left part we show the spectra of pions, kaons, antiprotons, and of the
$\phi$ mesons.  The parameters were fitted already in Ref. \cite{wbwf}
with the help of the spectra of pions, kaons and antiprotons only. The
good agreement of the measured $\phi$-meson spectrum \cite{yama} with
the model calculation supports our approach, since the interaction of
the $\phi$ meson with the hadronic enviroment is negligible, and
$\phi$ may be regarded as a good thermometer of the hadronic
system. The right part of Fig. 2 shows the prediction of our model for
the spectra of hyperons. The data for ${\bar \Lambda}$ come from
\cite{snellings}. Since these data are not normalized, we have
adjusted their normalization arbitrarily. \footnote{ The
newly-released normalized data on ${\bar \Lambda}$ production
\cite{bellwied} are consistent with our norm.}

\begin{figure}[t]
\epsfysize=6.1cm
\mbox{\epsfbox{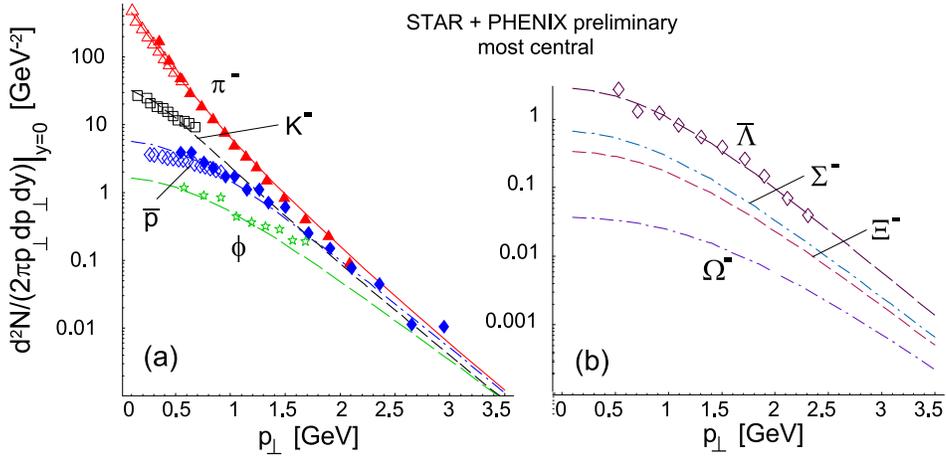}}
\label{mostc}
\caption{The $p_\perp$-spectra of various hadrons measured in
the most central collisions at RHIC ($\sqrt{s}$ = 130 GeV A). The
filled symbols are the data from PHENIX \cite{phenix}, the open
symbols are the STAR data [26-28] 
The curves represent the spectra obtained in the single-freeze-out
model. }
\end{figure}

Let us now make a few comments about the size of our geometric
parameters.  For the most central events, the fitted value of
$\rho_{\rm max} \sim$ 6 fm is very close to the radius of the
colliding gold nuclei, and is similar in magnitude with the measured
HBT transverse radius $R_s$.  The latter describes, however, an
average (r.m.s.)  transverse size, which is smaller than the true
geometric size of the system.  This means that our values of the
geometric parameters are too low.  This problem can be circumvented by
the inclusion of the excluded-volume corrections in the thermodynamic
description of the hadron gas \cite{pbmags,yg}. Such corrections make
the system more dilute, which can be compensated by an increase of
$\tau$ and $\rho_{\rm max}$ \cite{hirsch}. If we rescale $\tau$ and
$\rho_{\rm max}$ by the same factor, the spectra remain unchanged and
the agreement with the measured transverse radius can be achieved. On
the other hand, the time-extension of our system is much smaller than
the space-extension, so we expect that the experimental result
$R_{out}/R_{side} \approx 1$ will naturally appear in our model. The
calculation of the HBT radii in our model with the full treatment of
the hadronic resonances is in progress.

In conclusion, we emphasize that our approach explains in a very
economic way (4 parameters) both the relative hadron yields and the
$p_\perp$-spectra of all up-to-now measured hadrons
($\pi^+,\pi^-,K^+,K^-, p,{\bar p},{\bar \Lambda},\phi$). The model
includes in a transparent way the decays of the resonances, and the
longitudinal and transverse flow. Further extensions of the
application of the model should include the eliptic-flow effects and
the rapidity dependence.


\begin{thebibliography}{99}

\bibitem{pbmags}  P. Braun-Munzinger, J. Stachel, J. P. Wessels, and N. Xu,
Phys. Lett. {\bf B344}, 43 (1995); Phys. Lett. {\bf B365}, 1 (1996).

\bibitem{raf} J. Rafelski, J, Letessier, and A. Tounsi, Acta Phys. Pol. 
{\bf B28}, 2841 (1997).

\bibitem{cest}  J. Cleymans, D. Elliott, H. Satz, and R. L. Thews, Z. Phys.
{\bf C74}, 319 (1997).

\bibitem{pbmsps}  P. Braun-Munzinger, I. Heppe, and J. Stachel, Phys. Lett.
{\bf B465}, 15 (1999).

\bibitem{yg}  G. D. Yen and M. I. Gorenstein, Phys. Rev. {\bf C59}, 2788
(1999).

\bibitem{becatt} F. Becattini, J. Cleymans, A. Keranen, E. Suhonen,
and K. Redlich, Phys. Rev. {\bf C64}, 024901 (2001).

\bibitem{gaz} M. Ga\'zdzicki, Nucl. Phys. {\bf A681}, 153 (2001).

\bibitem{raf0} J. Rafelski, J. Letessier, and G. Torrieri,
Phys. Rev. {\bf C64}, 054907 (2001).

\bibitem{pbmrhic}  P. Braun-Munzinger, D. Magestro, K. Redlich, and J.
Stachel, Phys. Lett. {\bf B518}, 41 (2001).

\bibitem{fbm}  W. Florkowski, W. Broniowski, and M. Michalec,
Acta Phys. Pol. {\bf B33}, 761 (2002).

\bibitem{wbwf} W. Broniowski and W. Florkowski, Phys. Rev. Lett.  {\bf 87},
272302 (2001).

\bibitem{str} W. Broniowski and W. Florkowski, nucl-th/0112043. 

\bibitem{hirsch} W. Broniowski and W. Florkowski, to appear in the 
Proceedings of the Int. Workshop XXX on Gross Properties of Nuclei
and Nuclear Excitations, Hirschegg, 2002, hep-ph/0202059.

\bibitem{bjorken} J. D. Bjorken, Phys. Rev. {\bf D27}, 140 (1983).

\bibitem{baym} G. Baym, B. Friman, J.-P. Blaizot, M. Soyeur, and W. Czy\.z,
Nucl. Phys. {\bf A407}, 541 (1983). 

\bibitem{nikolaev} P. Milyutin and N. N. Nikolaev, Heavy Ion Phys {\bf 8}, 
333 (1998); V. Fortov, P. Milyutin, and N. N. Nikolaev,
JETP Lett. {\bf 68}, 191 (1998).

\bibitem{siemens} P. J. Siemens and J. Rasmussen, Phys. Rev. Lett. {\bf 42},
880 (1979); P. J. Siemens and J. I. Kapusta, Phys. Rev. Lett. {\bf 43},
1486 (1979).

\bibitem{schnedermann} E. Schnedermann, J. Sollfrank, and U. Heinz, 
Phys. Rev. {\bf C48}, 2462 (1993).

\bibitem{BL} T. Cs\"{o}rg\H{o} and B. L\"{o}rstad, Phys. Rev. {\bf C54}, 1390
(1996).

\bibitem{rischke} D. H. Rischke and M. Gyulassy, Nucl. Phys. {\bf A697}, 701 
(1996); Nucl. Phys. {\bf A608}, 479 (1996).

\bibitem{phenix} J. Velkovska, PHENIX Collaboration,  Nucl. Phys. 
{\bf A698}, 507 (2002).

\bibitem{phenixhbt} K. Adcox et al., PHENIX Collaboration, nucl-ex/0201008.

\bibitem{rapp} R. Rapp and E. V. Shuryak, to appear in the 
Proceedings of the Int. Workshop XXX on Gross Properties of Nuclei
and Nuclear Excitations, Hirschegg, 2002, nucl-th/0202059.

\bibitem{qm2001} contributions to the Proceedings of the Quark Matter
2001 Conference, Nucl. Phys. {\bf A698} (2002).

\bibitem{phenixoff} K. Adcox et al., PHENIX Collaboration, nucl-ex/0112006.

\bibitem{harris} J. Harris, STAR Collaboration, contribution
to QM2001.

\bibitem{yama} E. T. Yamamoto, STAR Collaboration,  hep-ph/0112017

\bibitem{snellings} R. Snellings, STAR Collaboration, hep-ph/0111437.

\bibitem{bellwied} R. Bellwied, STAR Collaboration, hep-ph/0112250.



\end{thebibliography}
\end{document}